\documentstyle[12pt]{article}
\begin{document}

\begin{center}
{\ {\Large {\bf A Model for Ordinary Levy Motion}} \\\vskip .25in {\bf A.V.
Chechkin and V.Yu. Gonchar} \\\vskip .25 in Institute for Theoretical
Physics \\National Science Center \\Kharkov Institute of Physics and
Technology,\\Akademicheskaya St.1, Kharkov 310108 Ukraine,\\and\\Institute
for Single Crystals\\National Academy of Sciences of Ukraine ,\\Lenin ave.
60, Kharkov 310001 Ukraine} \\\vskip .25 in {\bf Abstract. }
\end{center}

We propose a simple model based on the Gnedenko limit theorem for simulation
and studies of the ordinary Levy motion, that is, a random process, whose
increments are independent and distributed with a stable probability law. We
use the generalized structure function for characterizing anomalous
diffusion rate and propose to explore the modified Hurst method for
empirical rescaled range analysis. We also find that the structure function
being estimated from the ordinary Levy motion sample paths as well as the
(ordinary) Hurst method lead to spurious ''pseudo-Gaussian'' relations.

PACS number(s): 02.50.-r, 05.40.+j \newline

By Levy motions, or Levy processes, one designates a class of random
functions, which are a natural generalization of the Brownian motions, and
whose increments are stably distributed in the sense of P. Levy\cite{1}. Two
important subclasses are (i) ordinary Levy motions (oLm's), which generalize
the ordinary Brownian motion, or the Wiener process \cite{2}, and whose
increments are independent, and (ii) fractional Levy motions (fLm's), which
generalize the fractional Brownian motions (fBm's) \cite{3} and have an
infinite span of interdependence.

The Levy random processes play an important role in different areas of
applications, e.g., in economy \cite{4}, biology and physiology \cite{5},
fractal and multifractal analysis \cite{6}, problems of anomalous diffusion 
\cite{7} etc. In this paper we provide a simple method for numerical
simulation of oLm's and discuss their scaling properties.

At first we show the way to generate random sequence of independent
identically distributed (i.i.d.) random variables possessing stable
probability law. These variables play the role of increments of the oLm.

We restrict ourselves by symmetric stable law with the stable probability
density ${\it p_{\alpha ,D}(x)}$ and the characteristic function

\begin{equation}
\widehat{p}_{\alpha ,D}(k)=<e^{ikx}>=\exp \left( -D\left| k\right| ^\alpha
\right)  \label{1}
\end{equation}
Here $\alpha $ is the Levy index, $0<\alpha \leq 2,$ and {\it D }is a
positive parameter. At{\it \ $\alpha =1$ }and 2 one has the Cauchy and the
Gaussian probability laws, respectively. In other cases the symmetric stable
laws are not expressed in terms of elementary functions. At{\it \ $0<\alpha
<2$ }they have power law asymptotic tails{\it \ \cite{8}, 
\begin{equation}
p_{\alpha ,D}(x)\propto D\frac{\Gamma (1+\alpha )\sin (\pi \alpha /2)}{\pi
\left| x\right| ^{1+\alpha }},\quad x\rightarrow \pm \infty  \label{2}
\end{equation}
}Among the methods of generating random sequence with the given probability
law {\it F(x)} the method of inversion seems most simple and effective \cite
{9}. However, it is well-known fact that its effectiveness is limited by the
laws possessing analytic expressions for {\it F}$^{-1}$, hence, the direct
application of the method of inversion to the stable law is not expedient.
In this connection, we exploit an important property of stable
distributions. Namely, such distributions are limiting for those of properly
normalized sums of i.i.d. random variables \cite{10}. To be more concrete,
we generate the needed random sequence in two steps. At the first one we
generate an ''auxiliary'' sequence of i.i.d. random variables $\left\{ {\it %
\xi }_j\right\} $, whose distribution density {\it $F^{\prime }(x)${\ }}%
possesses asymptotics having the same power law dependence as the stable
density with the Levy index $\alpha $ has, see Eq.(2). However, contrary to
the stable law, the function $F(x)$ is chosen as simple as possible in order
to get analytic form of $F^{-1}.$ For example,\medskip

{\it 
\begin{equation}
F(x)=\bigskip \left\{ 
\begin{array}{l}
\left[ 2(1+\left| x\right| ^\alpha )\right] ^{-1}\quad x<0 \\ 
1-\left[ 2(1+x^\alpha )\right] ^{-1},\quad x\geq 0
\end{array}
\right.  \label{3}
\end{equation}
\strut }At the second step the normalized sum

{\it 
\begin{equation}
X(m)=\frac 1{am^{1/\alpha }}\sum_{j=1}^m\xi ,  \label{4}
\end{equation}
}where{\it \ 
\begin{equation}
\medskip a=\left( \frac \pi {2\Gamma (\alpha )\sin (\pi \alpha /2)}\right)
^{1/\alpha }  \label{5}
\end{equation}
\strut }is estimated. According to the Gnedenko theorem on the normal
attraction basin of the stable law \cite{10}, the distribution of the sum
(4) is then converges to the stable law with the characteristic function (1)
and {\it D }=1. It is reasonable to generate random variables having stable
distribution with the unit {\it D}, with a consequent rescaling, if
necessary. Repeating {\it N} times the above procedure, we get a sequence of
i.i.d. random variables{\it \ $\left\{ X_n(m)\right\} ,n=1,...,N$. }In the
top of Fig.1 the probability densities{\it \ ${p(x)}$ }for the members of
the sequence{\it \ $\left\{ X_n(m)\right\} (m=30)$ }are depicted by black
points for (a){\it \ $\alpha =1.0$, }and (b){\it \ $\alpha =1.5$. }The
functions{\it \ {p}$_{\alpha ,1}(x)$ }obtained with the inverse Fourier
transform, see Eq.(1), are shown by solid lines. In the bottom of Fig.1 the
black points depict asymptotics of the same probability densities in log-log
scale. The solid lines show the asymptotics given by Eq.(2). It is seen that
the Levy index can be estimated with the use of{\it \ $X_n$'}s, which lie
outside the peak located around{\it \ $x=0$. }The examples presented
demonstrate a good agreement between the probability densities for the
sequences {\it $\left\{ X_n\right\} $ }obtained with the use of the
numerical algorithm proposed and the densities of the stable laws.{\it \ }

We would like to stress that a certain merit of the proposed model is its
simplicity. It is entirely based on classical formulation of one of the
limit theorems and can be easily generalized for the case of asymmetric
stable distributions. It is also allows one, after some modifications, to
speed up the convergence to the stable law. These problems, however, ought
to be the subject of a separate paper. We note, that two schemes were
proposed recently, which use the combinations of random number generators 
\cite{11} and the family of chaotic dynamical systems with broad probability
distributions \cite{12}, respectively. The former method allows one to
generate the sequences with the symmetric laws, whereas the latter allows
one to generate also asymmetric ones. The comparison between our scheme and
those of Refs.\cite{11,12} is beyond the scope of our paper. In any case,
the proposed method can serve for a further constructing of non-stationary
processes and studying of their properties. We proceed to this task below.%
{\it \ }

With the help of the sequence obtained, the oLm is defined by

{\it \strut
\begin{equation}
L_\alpha (t)=\sum_{n=1}^tX_n  \label{6}
\end{equation}
\strut }(below we denote time argument as {\it t}, $\tau $ for the
continuous and for the discrete time scales as well;{\it \ {t}, $\tau $ }%
take positive integer values in the latter case).

In Fig.2 the stationary sequences of independent random variables obtained
with the numerical algorithm proposed are depicted by thin lines at 4
different Levy indexes. The thick lines depict the sample paths, or the
trajectories, of the oLm's. It is clearly seen that with the Levy index
decreasing, the amplitude of the increments increases. The large sparse
increments lead to large ``jumps'' (often named as ''Levy flights'') on the
trajectory.

{\it \strut }Let us proceed with the properties of self-similarity of the
oLm. The characteristic function of the oLm increments is{\it \ 
\begin{equation}
\left\langle \exp \left[ ik(L_\alpha (t+\tau )-L_\alpha (t)\right]
\right\rangle =\exp (-\left| k\right| ^\alpha \tau )  \label{7}
\end{equation}
($D=1$ }here and below). The increments of the oLm are stationary in a
narrow sense,{\it \ 
\begin{equation}
L_\alpha (t_1+\tau )-L_\alpha (t_2+\tau )\stackrel{d}{=}L_\alpha
(t_1)-L_\alpha (t_2)\quad ,  \label{8}
\end{equation}
}and self-similar with parameter 1/$\alpha $, that is, for an arbitrary{\it %
\ $h>0$ }

{\it 
\begin{equation}
L_\alpha (t+\tau )-L_\alpha (t)\stackrel{d}{=}\left\{ h^{-1/\alpha }\left[
L_\alpha (t+h\tau )-L_\alpha (t)\right] \right\} \quad ,  \label{9}
\end{equation}
}where{\it \ $\stackrel{d}{=}$ }implies that the two random functions have
the same distribution functions.

We consider two corollaries of Eqs.(7) - (9).

1. A ''1/$\alpha $ law'' for the generalized structure function (GSF) of the
oLm can be stated as follows: for all{\it \ $0<\mu <\alpha $ }the 1/{\it $%
\mu $}-th order root of the GSF is defined by{\it \ }

{\it \bigskip $\medskip $%
\begin{equation}
S_\mu ^{1/\mu }(\tau ,\alpha )=\left\langle \left| L_\alpha (t+\tau
)-L_\alpha (t)\right| ^\mu \right\rangle ^{1/\mu }=\tau ^{1/\alpha }V(\mu
;\alpha ),  \label{10}
\end{equation}
}where{\it \ 
\begin{equation}
V(\mu ;\alpha )=\left\{ \int\limits_{-\infty }^\infty dx_2\left| x_2\right|
^\mu \int\limits_{-\infty }^\infty \frac{dx_1}{2\pi }\exp (-ix_1x_2-\left|
x_1\right| ^\alpha )\right\} ^{1/\mu }\quad  \label{11}
\end{equation}
}For the ordinary Brownian motion $\alpha =2$, and 1/2 law is the indicator
of classical (normal) diffusion. Since $\tau $-dependence is not changed
with $\mu $ varying, then the quantity {\it $S_\mu ^{1/\mu }(\tau ;\alpha )$ 
}at any $\mu $ less than $\alpha ${\it \ }can serve as a measure of
anomalous diffusion rate. We remind that the (ordinary) structure function
is infinite for $\alpha <2$.

We study numerically the dependence of the index {\it s} in the relation{\it %
\ 
\begin{equation}
S_\mu ^{1/\mu }(\tau ;\alpha )\propto \tau ^s  \label{12}
\end{equation}
}vs $\mu ,\alpha .$ In Fig.3 $s$ vs $\alpha $ is depicted by crosses at
fixed $\mu =1/2.$ The 1/$\alpha $ curve is shown by primes. One can be
convinced himself that the 1/$\alpha $ law for the GSF is well confirmed at $%
\mu $ smaller than the smallest Levy index in numerical simulation. For the
comparison $s$ vs $\alpha $ is depicted by black points for the structure
function, that is, for $\mu =2$. It is shown that the structure function,
being estimated from a finite sample path, lead to the spurious
''pseudo-Gaussian'' value{\it \ $s=1/2$. }At the inset $s$ vs $\mu $ is
depicted for the oLm with $\alpha =1.$ It is shown that $s\cong 1$ at $\mu
\leq 1$ , whereas with $\mu $ increasing the deviation from 1/$\alpha $ law
increases.

The ''pseudo-Gaussian'' $\tau -$ dependence of the structure function can be
explained by the finiteness of sample length taken into account. Indeed, let 
$X_{\max }$ be the mode of maximum value (that is, the most probable maximum
value) for the sequence $\left\{ X_n\right\} ,$ which consists from $t$
terms having stable distribution with the Levy index $\alpha $. It can be
easily shown that $X_{\max }\propto t^{1/\alpha }$, and for the variance we
get $\left\langle X^2\right\rangle \propto t^{2/\alpha -1}$. Therefore, for $%
\tau $ smaller than $t$ (which is the natural condition when estimating the
structure function in numerical simulation or at data processing) one gets
in the discrete time scale {\it 
\begin{equation}
\left\langle \left( L_\alpha (t+\tau )-L_\alpha (t)\right) ^2\right\rangle
=\left\langle \left( \sum_{t+1}^{t+\tau }X_n\right) ^2\right\rangle \approx
\tau \left\langle X_t^2\right\rangle \propto \tau t^{2/\alpha -1}.
\label{13}
\end{equation}
}Thus, the square root of the structure function behaves as $\tau ^{1/2}$
for all $\alpha $'s, as it is indeed demonstrated in Fig.3. Furthermore, the
value of the structure function grows with the number $t$ of the terms in
the sample path growth. On the contrary, the GSD of the $\mu $-th order,
being estimated from a finite sample path at $\mu <\alpha $, does not grow
with $t$ growth.

2. A ''1/$\alpha $ law'' for the span of {\it L}$_\alpha (t)$ can be stated
as follows:

{\it 
\begin{equation}
R(\tau )=\sup_{0\leq {t}\leq \tau }L_\alpha (t+\tau )-\inf_{0\leq {t}\leq
\tau }L_\alpha (t) \stackrel{d}{=} \tau ^{1/\alpha }R(1)  \label{14}
\end{equation}
} For the ordinary Brownian motion $\tau ^{-1/2}R(\tau )$ has a distribution
independent of $\tau $\cite{13}. In the empirical rescaled range analysis,
that is, at experimental data processing or in numerical simulation the span
of the random process is divided by the standard deviation (that is, the
square root of the second moment) for the sequence of increments, which
''smoothes'' the variations of the span on the different segments of time
series \cite{14}. Such a procedure, called the Hurst method or the method of
normalized span, is not satisfactory for the Levy motion because of the
infinity of the theoretical value of the standard deviation. Therefore, we
propose to modify the Hurst method by exploiting the 1/$\alpha $ -th root of
the $\alpha $-th moment instead of standard deviation, that is, {\it 
\begin{equation}
\sigma _\alpha =\left( \frac 1\tau \sum_{n=1}^\tau \left| X_n\right| ^\alpha
\right) ^{1/\alpha }  \label{15}
\end{equation}
}

Since it has only weak logarithmic divergence with the number of terms in
the sum increasing, then one has{\it \ 
\begin{equation}
\overline{\left( \frac{R(\tau )}{\sigma _\alpha }\right) }\propto \tau
^H\quad ,  \label{16}
\end{equation}
}where $H\cong 1/\alpha $ is the Hurst index for the oLm with the Levy index 
$\alpha $, and the bar denotes averaging over the number of segments (having
the length $\tau $) of the sample path.

Fig.4 demonstrates the application of the modified Hurst method to the
sample path of the oLm with the Levy index $\alpha =1$. In Fig.4a the
fluctuations of span (thin curve) and those of $\sigma _\alpha $ (thick
curve) are shown for the case, when the total length of the sample is
divided into 64 segments, each of $\tau =16$ lengthwise. Below the
variations of the ratio $R/\sigma _\alpha $ are depicted. It is shown that
fluctuations of the ratio is much smaller than those of the span. This
circumstance justify the use of the ratio in the empirical analysis. In
Fig.4b the rescaled span vs time interval $\tau $ is depicted in log-log
scale by black points. The slope of the solid line is equal to $H=0.9$.

In Fig.5 the {\it H} vs $\alpha $ is depicted by crosses, whereas the curve
1/$\alpha $ is indicated by primes. The Hurst index obtained from a
''traditional'' ratio $R/\sigma _2$ is shown by black points. It follows
from the figure that $\sigma _\alpha $ only ''smoothes'' the variations of
span, thus leading to the correct value $H=1/\alpha $. On the contrary, the
standard deviation, being used in empirical analysis of the oLm,
''suppresses'' the variations of the span, thus giving rise to the spurious
value $H\approx 0.5$. As in case of the structure function, this
''pseudo-Gaussianity'' is explained by the finiteness of the sample of the
oLm. Indeed, for the segment of the length $\tau $ $R(\tau )\propto \tau
^{1/\alpha }$, whereas $\sigma _2\propto \tau ^{1/\alpha -1/2}$, and thus, $%
R/\sigma _2\propto \tau ^{1/2}$ for all $\alpha $'s. This circumstance
allows us to suggest that at estimating the (ordinary) structure function
and normalized span from experimental data the ''Levy nature'' of them can
be easily masked. This, in turn, poses an interesting task of developing
statistical methods for extracting reliable characteristics from
experimental data, for which Levy statistics can be expected from e.g., some
physical reasons.

At the end we note that introduction of correlations into the sequences of
i.i.d. stably distributed random variables allows us to extend the presented
methods on the studies of fractional Levy motions.{\it \ }

This work is done within the framework of the Project ''Chaos-2'' of the
National Academy of Sciences of Ukraine and the Project INTAS 93-1194. The
information support within the Project INTAS LA-96-09 is also acknowledged.

{\it \newpage \strut }FIGURE CAPTIONS

1. Probability densities (above) and their asymptotics (below) are indicated
for the sequences of random variables generated with the use of the proposed
numerical algorithm at the Levy indexes (a) $\alpha =1$ , and (b) $\alpha
=1.5$. The probability densities and the asymptotics of the stable laws are
indicated by solid lines.

2. Stationary sequences (thin lines) and ordinary Levy motion trajectories
(thick lines) at the different Levy indexes.

3. Plots of the exponent s in Eq.(12) versus the Levy index $\alpha $ at $%
\mu =1/2$ (crosses) and $\mu =2$ (black points). The 1/$\alpha $ curve is
depicted by dashed line. At the inset $s$ vs $\mu $ at $\alpha =1$ is shown.

4. (a) The variations of span (thin curve), of the GSD of the $\alpha $-th
order (thick curve) and of their ratio (below) at the different time
intervals for the oLm with $\alpha =1$. (b) Rescaled span vs time interval
in log-log scale (black points). Solid line has a slope $H=0.9${\it . }

5. Plots of the Hurst exponent {\it H} vs $\alpha $ estimated with the use
of Eq.(16) (crosses) and with the use of the ''traditional'' Hurst method
(black points). The 1/$\alpha $ curve is depicted by dashed line.

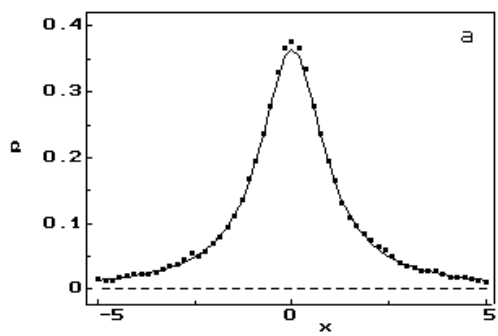
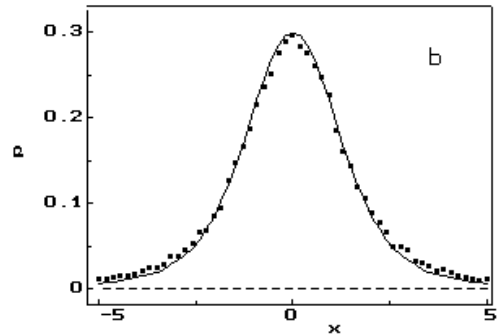
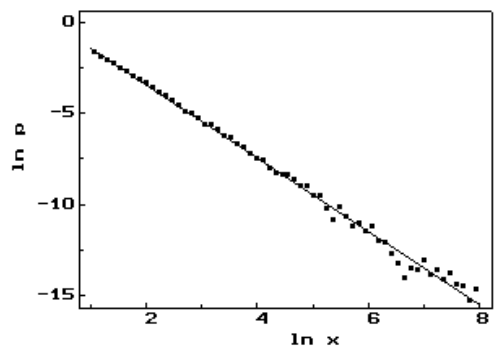
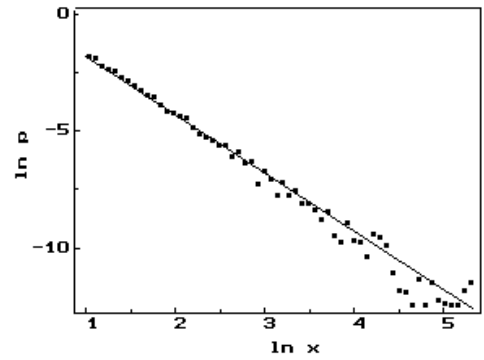
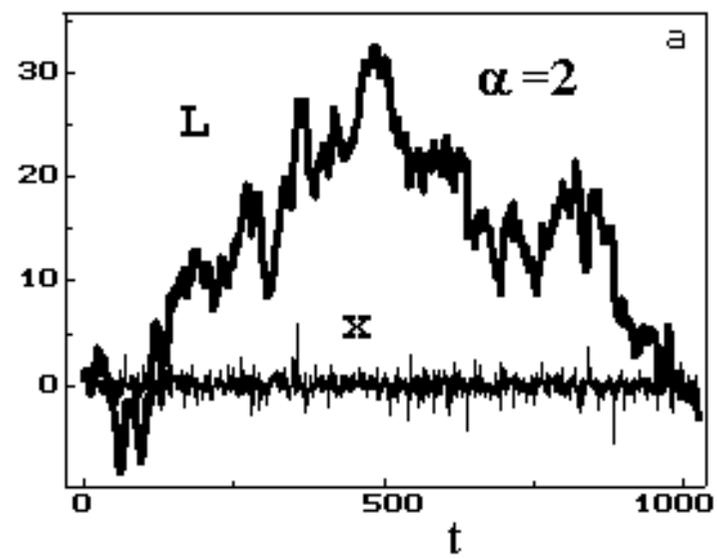
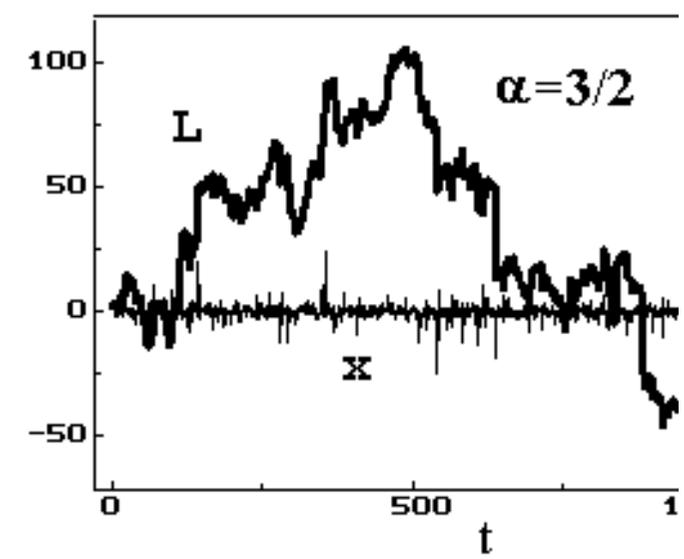
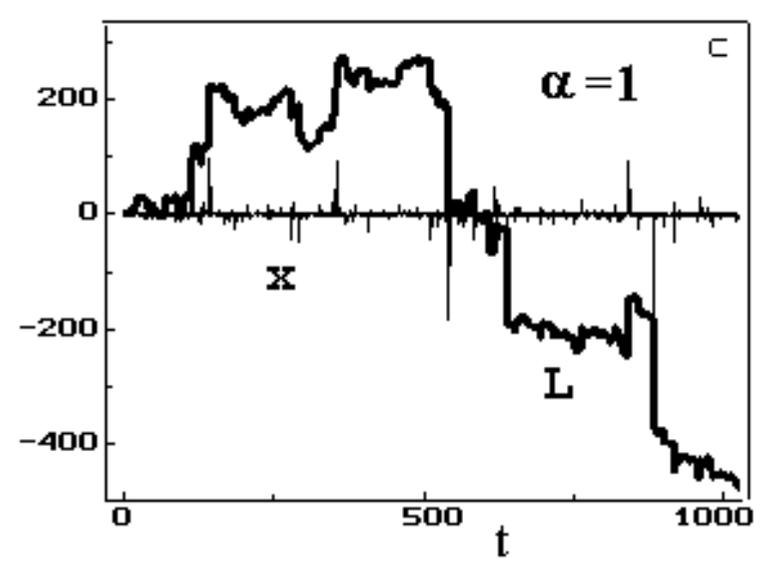
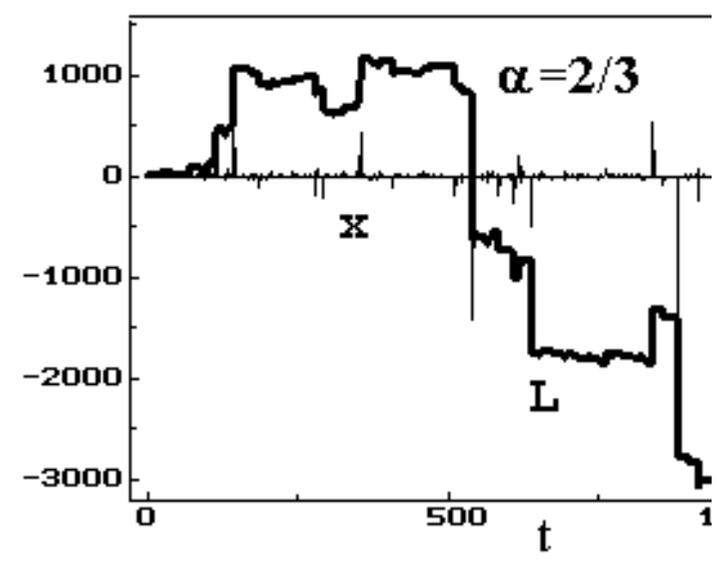
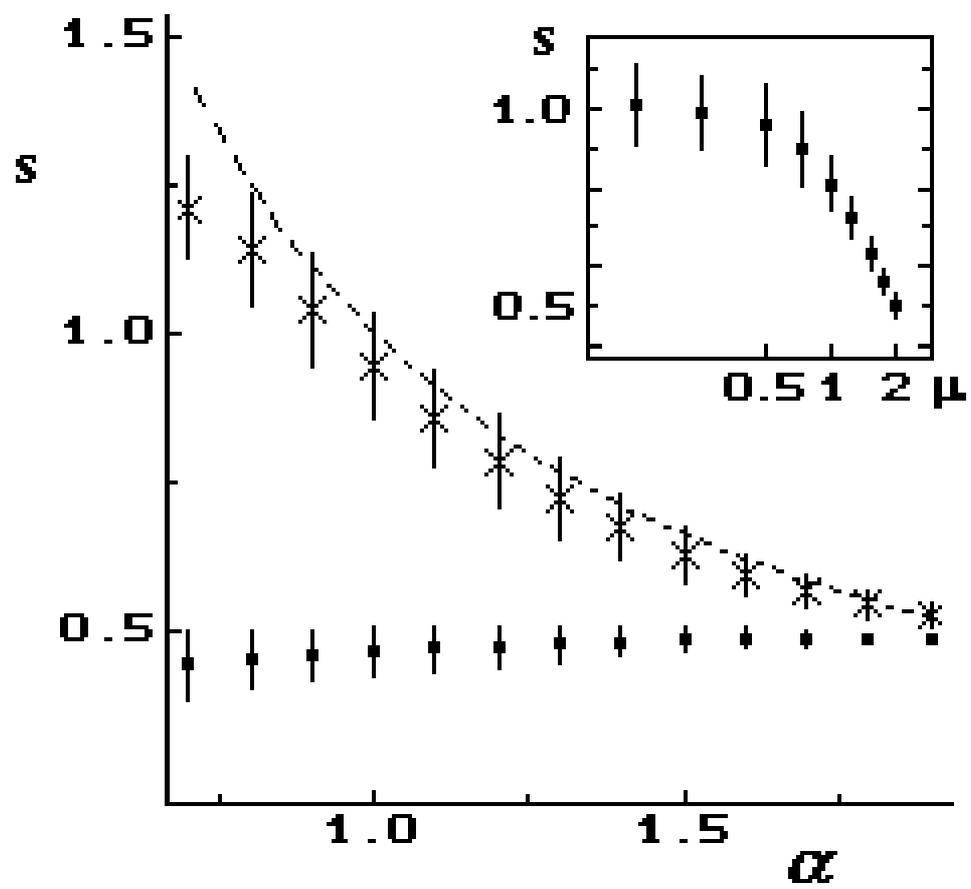
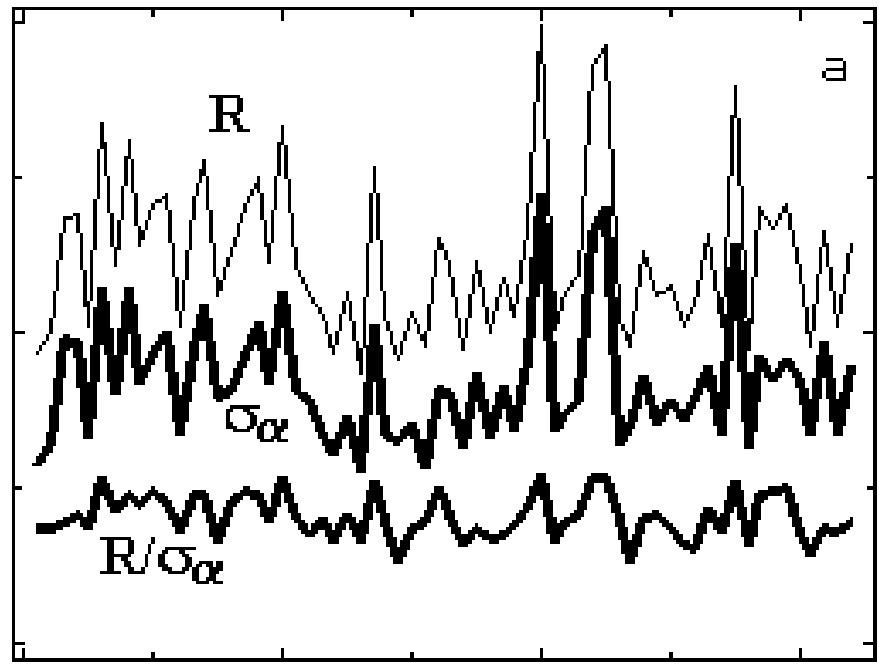
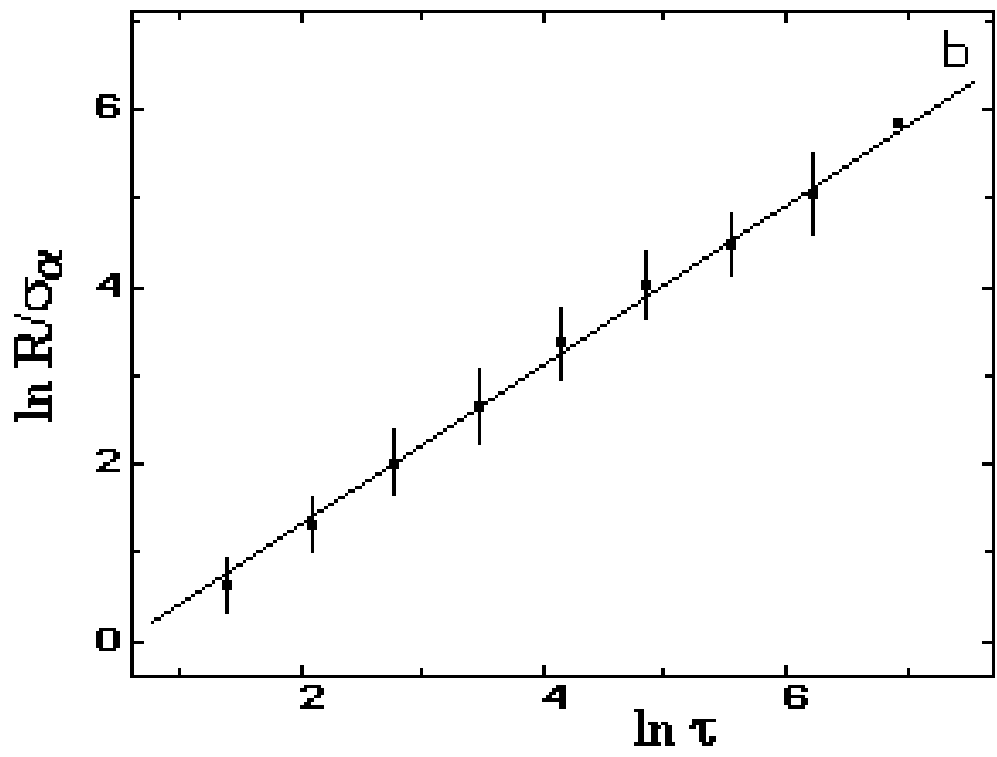
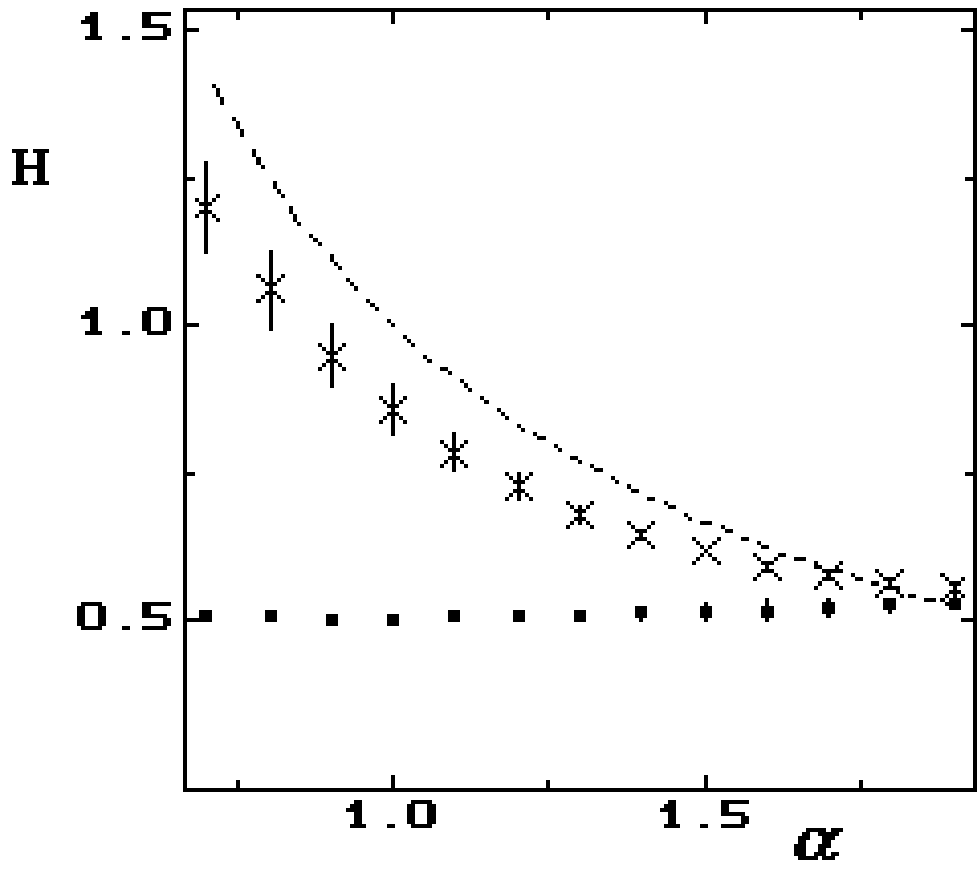
\end{document}